\documentclass[graybox]{svmult}

\usepackage{amsfonts}
\usepackage{amssymb}
\usepackage{amstext}

\usepackage{helvet}         
\usepackage{courier}        
\usepackage{type1cm}        
\usepackage{graphicx}        
\usepackage{multicol}        
\usepackage[bottom]{footmisc}

\def\be{\begin{equation}}
\def\ee{\end{equation}}
\def\ba{\begin{eqnarray}}
\def\ea{\end{eqnarray}}
\def\lb{\label}
\def\nn{\nonumber}

\def\a{\alpha}
\def\b{\beta}

\def\d{\delta}

\def\l{\lambda}

\def\E{{\cal E}}

\def\bU{\overline{U}_q}

\def\1{1\!\!{\rm I}}

\def\subbbc{{\rm C}\kern-3.3pt\hbox{\vrule height4.8pt width0.4pt}\,}

\def\vac{\mid \! 0 \rangle}

\def\Fd{{\cal F}^{diag}}
\def\Fp{{\cal F}'}
\def\bm{\mathbf m}

\def\Ba{\bar a}
\def\Bp{\bar p}

\begin{document}

\title*{"Spread" restricted Young diagrams from\\ a $2D$ WZNW dynamical quantum group}
\titlerunning{"Spread" restricted Young diagrams from a $2D$ WZNW dynamical QG}
\author{Ludmil Hadjiivanov and Paolo Furlan}
\institute{
Ludmil Hadjiivanov \at 
Elementary Particle Theory Laboratory,
Institute for Nuclear Research and Nuclear Energy (INRNE),
Bulgarian Academy of Sciences,
Tsarigradsko Chaussee 72, BG-1784 Sofia, Bulgaria,
\email{lhadji@inrne.bas.bg}
\and
Paolo Furlan \at Dipartimento di Fisica, Universit\`a degli Studi di Trieste, Strada Costiera 11, I-34014
Trieste, Italy,
\email{furlan@ts.infn.it}}
\maketitle

\abstract{
The Fock representation of the $Q$-operator algebra for the diagonal 2D
${\widehat{su}}(n)_k\,$ WZNW model where
$Q=(Q^i_j)\,,\ Q^i_j = a^i_\a \otimes \Ba^\a_j\,,$ and
$a^i_\a\,,\, \Ba^\b_j\,$ are the chiral WZNW "zero modes", has a natural basis labeled by $su(n)$ Young diagrams $Y_\bm\,$ subject to the  "spread" restriction
\newline\\
\centerline{\fbox {${\rm spr}\,(Y_\bm ):= \#$(columns) $+$ $\#$(rows) $\le k+n =: h\,.$} } }



\section{Introduction}
\label{intro}

This work contains a brief exposition of new results based on ideas and techniques, some parts of which have been already
made public in \cite{HF3, HF4, FHT5}. The latter relied, in turn, on the notion of quantum matrix algebras generated by
the chiral zero modes of the $SU(n)_k\,$ Wess-Zumino-Novikov-Witten (WZNW) model introduced in \cite{HIOPT}
(see also \cite{I2, FHIOPT, FHT6}). The relation of such algebraic objects with quantum groups \cite{D, J}
has been anticipated in \cite{AF, FG}. For generic values of the deformation parameter $q\,$ the Fock representation
of the chiral zero modes' algebra is a model space of $U_q(s\ell (n))\,$ \cite{BF, FHIOPT, FHT5}. In the most interesting
applications the deformation parameter $q\,$ is a root of unity (in our case we take $q = e^{-i\frac{\pi}{h}}\,,\ h= k+n$).
It has been shown, in particular, in \cite{FHT7} that the Fock representation of the chiral zero modes' algebra for $n=2\,$
carries a representation of the {\em restricted} (finite dimensional) quantum group $\bU(s\ell (2))\,$ of \cite{FGST1, FGST2}
containing, as submodules or quotient modules, all irreducible representations of the latter.

Combining the left and right chiral zero modes' algebras, one obtains a particular ($2D$ zero modes')
{\em dynamical} quantum group \cite{EV2}. Its role in the description of the internal (sector) structure  of the $n=2\,$
WZNW model has been studied in \cite{FHT2, DT} where it has been shown that it provides in a natural way a finite extension
of the unitary model. The setting is reminiscent to the axiomatic (cohomological) approach to gauge theories,
the quantum group playing the role of generalized gauge symmetry.

The main statement of the present paper is that the Fock representation of the $2D$ zero modes' algebra has a basis
that is in a one-to-one correspondence with the finite set of $su(n)\,$ Young diagrams \cite{Ful} restricted by a
specific "spread" condition. These fit into a rectangle of size $(n-1)\times (h-1)\,$ which is thus wider than the
$(n-1)\times k\,$ rectangle containing the unitary $\widehat{su}(n)_k\,$ fusion sectors \cite{DFMS}.
(The ring structure of the latter, i.e. the Verlinde algebra, is conveniently described by a suitable representation
of the phase model hopping operator algebra or, in other terms, of the affine local plactic algebra \cite{KS, Wa2012}.)
Note that the spread restriction is more stringent than just the fitting into the $(n-1)\times (h-1)\,$ rectangle requirement.

It would be interesting to find out if the affine algebra representations (some of which non-integrable) corresponding to
the finite set of "spread restricted" $su(n)\,$ diagrams constitute a sensible extension of the unitary WZNW model.
We hope that, on the long run, the present approach could help to better understand the adequate "gauge" symmetry
(the $2D\,$ counterpart of the Doplicher-Roberts \cite{DR} compact group) governing the "addition of non-abelian charges",
i.e. the fusion rules, in RCFT (cf. \cite{BNS, Hayashi, NV00, BK, EN, PZ, ENO}).

\section{Definitions: $SU(n)_k\,$ WZNW zero modes}
\label{sec2}

We will recall here the basic assumptions about the chiral and $2D\,$ WZNW "zero modes" and their Fock representation, justified by the consistent application of the principles of canonical quantization (see e.g. \cite{HF4, FHT5}).

The mutually commuting left and right $SU(n)_k\,$ WZNW chiral zero modes' algebras ${\cal M}_q\,,\,{\bar{\cal M}}_q\,$ are generated by operators $\{q^{p_j}\,, a^i_\a\}\,$ and $\{q^{{\bar p}_j}\,, \Ba_i^\a\}\,,$ respectively, satisfying identical exchange relations:
\ba
&&q^{p_i} q^{p_j} = q^{p_j} q^{p_i}\ ,\quad \prod_{j=1}^n q^{p_j} = 1\ ,\quad
q^{p_{j\ell}} a_{\alpha}^i = a_{\alpha}^i \, q^{p_{j\ell} + \delta_j^i - \delta_{\ell}^i}\quad ( p_{j\ell}:= p_j-p_\ell )\ ,\nn\\
&&q^{{\bar p}_i} q^{{\bar p}_j} = q^{{\bar p}_j} q^{{\bar p}_i}\ ,\quad \prod_{j=1}^n q^{{\bar p}_j} = 1\ ,\quad
q^{\bar p_{j\ell}}\, \bar a^\a_i = \bar a^\a_i\, q^{\bar p_{j\ell} + \d_{ij} - \d_{i \ell}}\quad ( {\bar p}_{j\ell}:= {\bar p}_j-{\bar p}_\ell)\ ;\qquad
\lb{pa}
\ea
all indices run from $1\,$ to $n\,.$ Bilinear combinations of chiral zero modes intertwine dynamical and constant $R$-matrices;
the left and right sector quadratic exchange relations following from
\be
\hat R_{12}(p) \, a_1 \, a_2 = a_1 \, a_2 \, \hat R_{12}\ ,
\quad
\hat R_{12}\, \bar a_1\, \bar a_2\, =\, \bar a_1\, \bar a_2\, \hat{\bar R}_{12}(\bar p)\quad
(\, \hat{\bar R}_{12}(\bar p) = ({\hat R}_{12} (\bar p))^t\, )\ ,
\lb{ExR}
\ee
respectively, coincide as well when written in components:
\ba
&&a^j_\b a^i_\a\, [p_{ij}-1] = a^i_\a a^j_\b\, [p_{ij}] -\,a^i_\b\, a^j_\a \, q^{{\epsilon}_{\a\b}p_{ij}} \quad
(\,{\rm for}\quad i\ne j \quad {\rm and}\quad\alpha\ne\beta )\ ,\nn\\
&&[a^j_\alpha , a^i_\alpha ] = 0\ ,\qquad a^i_\alpha a^i_\beta = q^{{\epsilon}_{\alpha\beta}}\, a^i_\beta a^i_\alpha\ ,
\lb{aa2}\\
&&\Ba_j^\b \Ba_i^\a\,[\Bp_{ij}-1]
= \Ba_i^\a \Ba_j^\b\,[\Bp_{ij}] - \,\Ba_i^\b\, \Ba_j^\a \,q^{{\epsilon}_{\a\b}\Bp_{ij}} \quad
(\,{\rm for}\quad i\ne j \quad {\rm and}\quad\alpha\ne\beta\, )\ ,\nn\\
&&[\Ba_j^\alpha , \Ba_i^\alpha ] = 0\ ,\qquad \Ba_i^\alpha \Ba_i^\beta =
q^{{\epsilon}_{\alpha\beta}}\, \Ba_i^\beta \Ba_i^\alpha
\lb{aa2barn}
\ea
(the antisymmetric symbol $\epsilon_{\a\b} = \pm 1\,$ for $\a \gtrless \b\,$ and vanishes for $\a=\b$).
They are supplemented by appropriate $n$-linear determinant conditions,
\be
\lb{Dq}
\det (a) = {\cal D}_q(p) \ ,\qquad \det (\Ba) = {\cal D}_q (\bar p)\ ,
\ee
where ${\cal D}_q(p) := \prod_{i<j} [p_{ij}]\,$ and
\ba
&&\det (a) := \frac{1}{[n]!} \, \epsilon_{i_1 \ldots i_n} \, a_{\alpha_1}^{i_1} \ldots a_{\alpha_n}^{i_n} \,
\varepsilon^{\alpha_1 \ldots \alpha_n}\ ,\nn\\
&&\det (\Ba) := \frac{1}{[n]!} \, \varepsilon_{\alpha_1 \ldots \alpha_n}\,\Ba_{\alpha_1}^{i_1} \ldots \Ba_{\alpha_n}^{i_n} \,
\epsilon^{i_1 \ldots i_n}\ ,
\lb{DqaBa}
\ea
$\epsilon_{i_1 \ldots i_n} = \epsilon^{i_1 \ldots i_n}\,$ and $\varepsilon_{\alpha_1 \ldots \alpha_n} = \varepsilon^{\alpha_1 \ldots \alpha_n}\,$
being the "ordinary" and "quantum" fully ($q$-)antisymmetric $n$-tensors, respectively. Finally, for $q^h=-1\,$ the chiral zero modes' algebras
${\cal M}_q\,,\,{\bar{\cal M}}_q\,$ possess non-trivial two-sided ideals such that the corresponding factor algebras ${\cal M}^{(h)}_q\,$ and ${\bar{\cal M}}^{(h)}_q\,$ are characterized by the additional relations
\be
(a^i_\a)^h = 0 \ ,\quad q^{2h\,p_{j\ell}} = 1\ ,\qquad
(\Ba^\a_i)^h = 0\ ,\quad q^{2h\,{\bar p}_{j\ell}} = 1\ .
\lb{ah0}
\ee
(Strictly speaking, the two algebras are identified with the corresponding {\em non-commutative polynomial} rings in $a^i_\a\,$ and $\Ba^\a_i\,$
over the fields of rational functions of $q^{p_j}\,$ and $q^{{\bar p}_j}\,,$ respectively.) We will be interested in the Fock space representation ${\cal F}^{(h)} \otimes {\bar{\cal F}}^{(h)}\,$ of ${\cal M}^{(h)}_q\otimes{\bar{\cal M}}^{(h)}_q\,,$ where
\be
{\cal F}^{(h)} = {\cal M}^{(h)}_q \vac\ ,\qquad {\bar{\cal F}}^{(h)} = {\bar{\cal M}}^{(h)}_q \vac\ .
\lb{Fock-h2}
\ee
The action of the generating elements on the vacuum vector is subject to
\ba
&&q^{p_{j\ell}}\vac = q^{\ell -j}\vac = q^{\bar p_{j\ell}}\vac \ ,\qquad j,\ell = 1,\dots , n\ ,\nn\\
&&a^i_\a \vac = 0 = \Ba^\a_i \vac\ ,\qquad i = 2,\dots , n\ .
\lb{Mqvac}
\ea
It follows that monomials in $a^i_\a\,,\, \Ba^\a_i\,$ generate eigenvectors of $q^{p_j}\,$ and $q^{{\bar p}_j}\,,$ respectively,
with eigenvalues of $p_{j\ell}\,,\,{\bar p}_{j\ell}\,$ corresponding to {\em shifted} $su(n)\,$ weights (e.g. $p_{jj+1} = \l_j+1\,,$ the vacuum quantum numbers being given by the components of the Weyl vector).

Defining $Q^i_j = a^i_\a \otimes \Ba^\a_j \in {\cal M}^{(h)}_q\otimes{\bar{\cal M}}^{(h)}_q\,,$ we will call the corresponding operator algebra "the $Q$-algebra" (of the $SU(n)_k\,$ WZNW model). Our task will be to describe the structure of its vacuum representation as a subspace of the "extended" (carrier) space ${\cal F}^{(h)} \otimes {\bar{\cal F}}^{(h)}\,.$

This has been done in a completely satisfactory way for $n=2\,$ (in \cite{HF4, FHT5}; see also \cite{FHT2}) and the emerging picture is easy to describe. It turns out that in this case the diagonal elements of the matrix $Q = (Q^i_j)\,$ commute with the off-diagonal ones and both generate two copies of the (finite dimensional) {\em restricted quantum group} $\bU(s\ell(2))\,$ of \cite{FGST1, FGST2, FHT7}. The corresponding Fock space representations are however quite different: while the one generated by the off-diagonal elements of $Q\,$ is one dimensional, the diagonal $Q$-operators span a subspace ${\cal F}^{diag}\,$ of dimension
$h = k+2\,$ in the $h^4$-dimensional ${\cal F}^{(h)} \otimes {\bar{\cal F}}^{(h)}\,.$ Furthermore, there is a natural scalar product on ${\cal F}^{diag}\,$ which is positive semidefinite, the subspace of zero-norm vectors ${\cal F}''\,$ being one-dimensional, ${\cal F}'' =  {\mathbb C}\, (Q^1_1)^{h-1} \vac\,.$ One obtains in effect a finite dimensional toy generalization of axiomatic gauge theory, the role of the pre-physical subspace ${\cal F}'\,$ being played by ${\cal F}^{diag}\,$ and such that the physical subquotient
\be
{\cal F}^{phys} = {\cal F}' / {\cal F}'' \simeq \oplus_{p=1}^{h-1} {\cal F}^{phys}_p\ ,\qquad
{\cal F}^{phys}_p := {\mathbb C}\, (Q^1_1)^{p-1} \vac
\lb{Fph}
\ee
contains exactly the fusion sectors ${\cal F}^{phys}_p\ ( p=2I+1 )\,$ of the unitary model.

It is this picture that we would like to generalize to $n\ge 3\,$ when $q=e^{-i\frac{\pi}{h}}\,,\ h=k+n\,.$

\section{The $Q$-algebra for $n\ge 3$ and the space ${\cal F}'$}
\label{sec3}

For the lack of space we will only sketch in this section the derivation of the results for $n\ge 3\,$ postponing most of the interesting details to a forthcoming publication. First of all, it is easy to see 
that (\ref{aa2}), (\ref{aa2barn}) and (\ref{ah0}) imply
\be
(Q^i_j)^h = 0\ .
\lb{Qh0}
\ee
Combining further the quadratic exchange relations for the left and right sector zero modes (\ref{ExR}), we obtain those for the $Q$-operators in a {\em dynamical quantum group} form \cite{EV2, I2}:
\be
{\hat R}_{12} (p)\,Q_1\,Q_2 = Q_1\,Q_2\,{\hat {\bar R}}_{12}(\bar p)\ .
\lb{RpQQ}
\ee
(As in the case of chiral exchange relations (\ref{aa2}), (\ref{aa2barn}), we will actually postulate relations obtained after getting rid of the denominators in the entries of the two dynamical $R$-matrices.) A straightforward computation shows that, as a result, any two entries of the matrix $Q\,$ belonging to the same row or column commute.

All this has been already proved in \cite{HF4} where the problem of commutativity of diagonal and off-diagonal elements has been also addressed. The novelty we would like to announce here answers this question and also describes the $n\ge 3\,$ counterpart of the pre-physical space ${\cal F}'\,$ of Section 2, providing a basis in it labeled by (a certain finite set of) $su(n)\,$ Young diagrams.

To this end we first introduce again the space ${\cal F}^{diag}\,$ generated from the vacuum by diagonal $Q$-operators. Due to (\ref{Mqvac}),
the vacuum is annihilated by all $Q^i_j\,$ except for $i=j=1\,.$ Depicting the action of each diagonal operator $Q^j_j\,$ by adding a box to the $j$-th row of a table with $n\,$ rows, we can make correspond to any vector in
${\cal F}^{diag}\,$ generated by a monomial a unique tableau with boxes numbered in the order of appearance in the product (counted from the right) of the specific operator.

It is not clear from the outset even if ${\cal F}^{diag}\,$ is finite dimensional. However, one immediately realizes that any {\em single row table} containing more than $h-1\,$ boxes should vanish, due to (\ref{Qh0}) implying $v_h^{(1)} := (Q^1_1)^h \vac = 0\,.$ Consider next the vector $v_h^{(2)} := Q^2_2\, (Q^1_1)^{h-1} \vac\,.$ Noting that
on ${\cal F}^{diag}\,$ the eigenvalues of $p_{j\ell}\,$ and ${\bar p}_{j\ell}\,$ coincide and that (\ref{RpQQ}) implies, for any
$v \in {\cal F}^{diag}\,,$
\be
Q^i_j\, v = 0 \quad {\rm or}\quad Q^j_i\, v = 0 \quad \Rightarrow\quad [p_{ij}+1]\, Q^i_i Q^j_j\, v = [p_{ij}-1]\, Q^j_j Q^i_i \, v
\lb{pQpQ}
\ee
(see \cite{HF4}), we deduce from (\ref{pa}) and (\ref{Mqvac}) that (since $[h-2]=[2]\ne 0$)
\ba
&&[p_{21}+1] \,Q^2_2\, (Q^1_1)^{h-1} \vac
= [p_{21}-1] \,Q^1_1\, Q^2_2\, (Q^1_1)^{h-2} \vac\ ,\quad {\rm i.e.}\nn\\
&&-[h-2]\,v_h^{(2)} = -[h]\,w_h^{(2)} \equiv 0\quad \Rightarrow\quad v_h^{(2)} = 0\ ,
\lb{v20}
\ea
where $w_h^{(2)}:= Q^1_1\, Q^2_2\, (Q^1_1)^{h-2} \vac\,.$
Similar observations suggest to introduce the space
$\Fp \subseteq \Fd\,$ as the linear span of vectors of the type
\be
v_\bm := (Q^i_i)^{m_i}\dots (Q^2_2)^{m_2} (Q^1_1)^{m_1}\! \vac\ ,\quad \bm = (m_1, m_2, \dots , m_i, 0,\dots, 0)
\lb{vi}
\ee
with
\be
1\le i\le n-1\ ,\quad m_i\le m_{i-1} \le \dots \le m_1\ ,\quad
m_1 + i \le h\ .
\lb{Ym}
\ee
It turns out that it has remarkable properties.
\begin{itemize}
\item
$\Fp\,$ is finite dimensional
\item
it is annihilated by any off-diagonal $Q$-operator: $\ Q^j_\ell\, \Fp = 0$
\item
the basis (\ref{vi}) can be labeled by admissible $su(n)\,$ {\em Young diagrams} $Y_{\bm}\,$ of {\em maximal hook length} not exceeding $h-1\,$ or, which is the same, of {\em spread}
\be
{\rm spr}\, (Y_{\bm}) := i+ m_1 \le h\ .
\lb{spread}
\ee
\end{itemize}
The last assertion needs some clarification. We first verify, by using (\ref{pQpQ}), that reordering the factors in the $Q$-monomial (\ref{vi}) reproduces one and the same vector, up to a non-zero coefficient, as far the "standard $su(n)$ rule" $m_j \le m_{j-1}\,$ is respected at any step
(i.e., also in subdiagrams obtained by removing an arbitrary number of
$Q$-operators from the left); so under this condition the numeration of  boxes is irrelevant. What we call "maximal hook length" (of a non-trivial Young diagram) is the hook length of the box in the upper left (NW) corner; we have used the term "spread" for the sum of the numbers of columns and rows.

To show that $\Fp = \Fd\,,$ we have to prove that any action
violating the conditions (\ref{Ym}), or equivalently -- in the "diagrammatic language" -- that
\begin{itemize}
\item
adding a box to the $n$-th row
\item
adding an extra box to the $j$-th row when $m_j=m_{j-1}\,$ or, finally,
\item
adding an extra box to the first column or row of a diagram saturating the spread inequality (\ref{spread}), i.e. for which
${\rm spr}\, (Y_{\bm}) = h$
\end{itemize}
all lead to a zero vector (or to another vector in $\Fp$). The proof relies essentially on the careful application of (\ref{pQpQ}) (and, for the first point, also on the determinant conditions (\ref{Dq})). The only difficulty arises when one violates the spread inequality by adding an extra box to the first row of the diagram (i.e., $m_1 \to m_1+ 1\,$ for $i+m_1=h$).

In this last case the problem can be reduced to hook shaped diagrams (all boxes in a diagram saturating (\ref{spread}) become irrelevant except those in the first row and the first column which form its "backbone").
In effect, for $i > 2\,$ we just have to generalize (\ref{v20}), introducing
\be
v_h^{(i)} = Q^i_i Q^1_1 \, v\ , \quad w_h^{(i)} = Q^1_1 Q^i_i \, v\ ,\quad
v := Q^{i-1}_{i-1} \dots Q^2_2 \,(Q^1_1)^{h-i} \vac\ .
\lb{vi0}
\ee
The argument why $v_h^{(i)} = 0\,$ is the same as in (\ref{v20}); we only have to evaluate $[p_{i1}\pm 1] \, v = [1-h\pm 1]\, v\,.$ The problem is to show that $w_h^{(i)} = 0\,$ too. To tackle it, we need a new technique which is introduced in the next section.

Before going to it we will make the following important remark.
It is obvious that the $su(n)\,$ Young diagrams of spread restricted by
$h\,$ fit into a rectangle of size $(n-1)\times (h-1)\,.$ (The spread condition imposes a stronger restriction, except for $n=2$.)
In the WZNW setting the zero modes are coupled to elementary chiral vertex operators (CVO) with similar intertwining properties
\cite{FHT5}; having this in mind, we note that the {\em integrable} representations of the affine algebra $\widehat{su}(n)_k\,$
(or the fusion sectors of the unitary model) are labeled by {\em all} $su(n)\,$ Young diagrams that fit into the
"narrower" rectangle of size $(n-1)\times k\,.$ Extending the analogy with the $n=2\,$ case, cf. (\ref{Fph}), we would expect, in particular, the vectors $v_\bm\,$ (\ref{vi}) to have zero norm iff
they correspond to diagrams outside the "unitary" rectangle.
(All such diagrams have thus also boxes in the additional $(n-1)\times (n-1)\,$ square.)

The following figures illustrate the above ideas and notions.

\newpage

\begin{figure}[htb]
\centering
\includegraphics[width=1.17\textwidth]{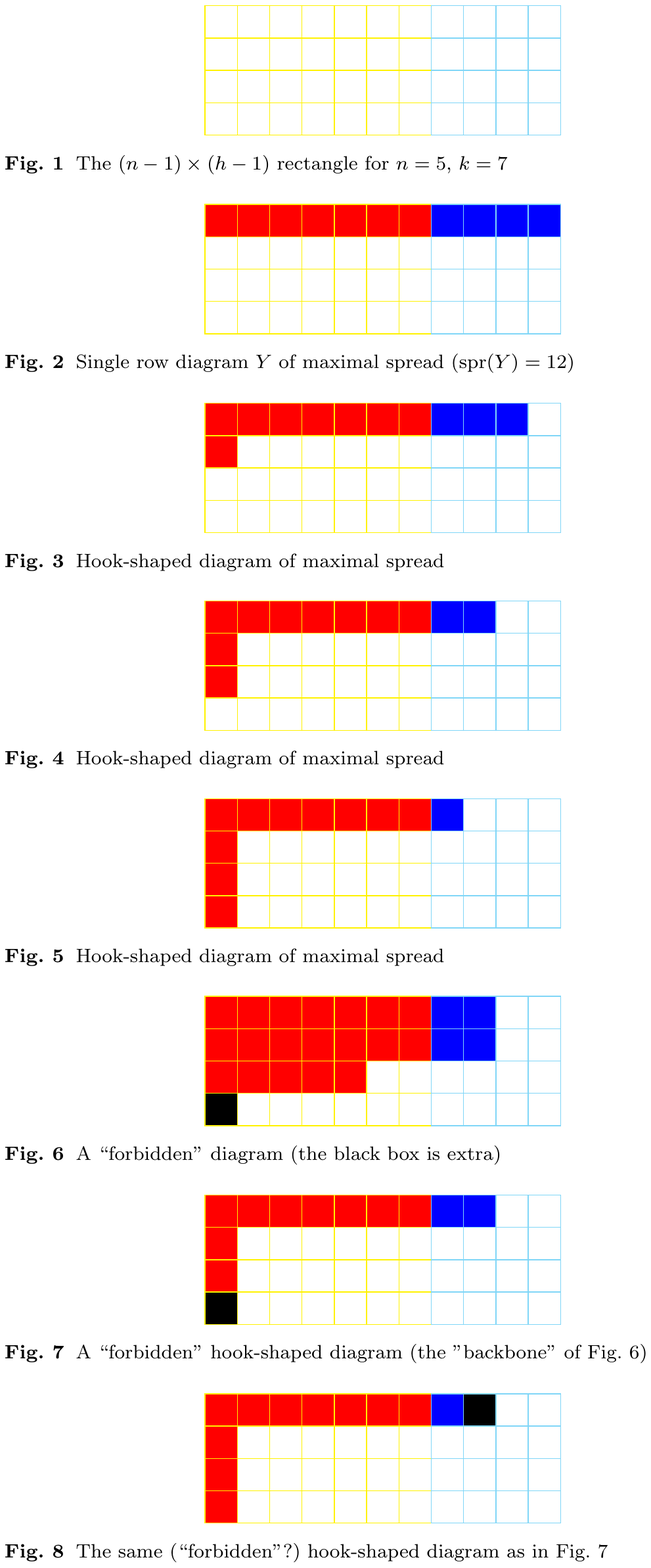}
\label{Diagrams-n}
\end{figure}

\section{Chiral $q$-symmetric and $q$-antisymmetric bilinears}
\label{sec4}

It turns out that the cumbersome bilinear exchange relations (\ref{aa2}), (\ref{aa2barn}) assume an amazingly simple form when written in terms of the corresponding $q$-antisymmetric and $q$-symmetric bilinear combinations
\be
a^i_\a a^j_\b = A^{ij}_{\a\b} + S^{ij}_{\a\b}\ ,\quad A^{ij}_{\a\b} = - q^{-\epsilon_{\a\b}} A^{ij}_{\b\a}\ ,
\quad S^{ij}_{\a\b} = q^{\epsilon_{\a\b}} S^{ij}_{\b\a}
\lb{SA}
\ee
defined by
\be
\ \ \, [2]\, A^{ij}_{\a\b} := a^i_{\a'} a^j_{\b'} A^{\a' \b'}_{~\a\b} = \left\{
\begin{array}{ll}
&q^{-\epsilon_{\a\b}} a^i_\a a^j_\b - a^i_\b a^j_\a\ ,\quad \a\ne \b \\
&0\ ,\quad \a=\b
\end{array}
\right.
\lb{Adef}
\ee
and
\be
[2]\, S^{ij}_{\a\b} := a^i_{\a'} a^j_{\b'} S^{\a' \b'}_{~\a\b} = \left\{
\begin{array}{ll}
&q^{\epsilon_{\a\b}} a^i_\a a^j_\b + a^i_\b a^j_\a\ ,\quad \a\ne \b \\
&[2]\, a^i_\a a^j_\a\ \ ( \equiv [2]\, a^j_\a a^i_\a)\ ,\quad \a=\b
\end{array}
\right.\ ,\quad
\lb{Sdef}
\ee
respectively, and their bar analogs
\be
[2]\, \bar A_{ij}^{\a\b} := A^{\a \b}_{~\a'\b'}\, \Ba_i^{\a'} \Ba_j^{\b'} \ ,\qquad
[2]\, \bar S_{{ij}}^{\a\b} := S^{\a \b}_{~\a'\b'}\, \Ba_i^{\a'} \Ba_j^{\b'}\ .
\lb{ASB}
\ee
A simple calculation \cite{FHT5} shows that relations (\ref{aa2}), (\ref{aa2barn}) for $i\ne j\,$ and $\a \ne \b\,$ are equivalent to
\ba
&&[p_{ij}+1]\, A^{ij}_{\a\b} = - [p_{ij}-1]\, A^{ji}_{\a\b} \ ,\qquad  S^{ij}_{\a\b}  = S^{ji}_{\a\b}  \ ,\nn\\
&&[\Bp_{{ij}}+1]\, {\bar A}_{{ij}}^{\a\b}  = - [\Bp_{{ij}}-1]\, {\bar A}_{{ji}}^{\a\b}\ ,\qquad
{\bar S}_{{ij}}^{\a\b} = {\bar S}_{{ji}}^{\a\b}\ .
\lb{AS-and-bar1}
\ea
The remaining relations (\ref{aa2}), (\ref{aa2barn}) look equally simple in these terms:
\be
 S^{ij}_{\a\a}  = S^{ji}_{\a\a} \ ,\quad A^{ii}_{\a\b} = 0\ , \qquad
{\bar S}_{{ij}}^{\a\a} = {\bar S}_{{ji}}^{\a\a} \ ,\quad { \bar A}_{ii}^{\a\b} = 0\ .
\lb{AS-and-bar2}
\ee
Note that
\be
S^{ij}_{\a\b}\otimes {\bar A}_{\ell m}^{\a\b} = 0 = A^{ij}_{\a\b}\otimes {\bar S}_{\ell m}^{\a\b}
\lb{SA=0a}
\ee
(where summation over $\a\,$ and $\b\,$ is assumed) since e.g.
\be
S^{ij}_{\a\b}\otimes {\bar A}_{\ell m}^{\a\b} = (q^{\epsilon_{\a\b}} S^{ij}_{\b\a}) \otimes
(- q^{-\epsilon_{\a\b}} {\bar A}_{\ell m}^{\b\a}) = - \,S^{ij}_{\b\a} \otimes {\bar A}_{\ell m}^{\b\a}\ \ ,
\lb{SabAab=0a}
\ee
and hence
\be
Q^i_\ell\, Q^j_m = (S^{ij}_{\a\b}+A^{ij}_{\a\b})\otimes ({\bar S}_{\ell m}^{\a\b}+{\bar A}_{\ell m}^{\a\b}) =
S^{ij}_{\a\b}\otimes {\bar S}_{\ell m}^{\a\b} + A^{ij}_{\a\b}\otimes {\bar A}_{\ell m}^{\a\b} \ .
\lb{QQSAa}
\ee
These relations turn out to be crucial as they shed light on the quantum group "internal structure" of monomials in the
$Q$-operators. (Their importance can be anticipated in the derivation of (\ref{Qh0}), see \cite{HF4}, which is actually based
on the $q$-symmetry underlying the chiral expansion of $Q^i_j Q^i_j$.) To illustrate the idea, we will demonstrate that,
presented as in (\ref{vi0}), the vector $w_h^{(i)}\,$ only contains the $q$-antisymmetric part of $Q^1_1 Q^i_i\,.$
Indeed, it follows from $[p_{i1}-1]\,v=0\,$ and (\ref{AS-and-bar1}) that
\be
[p_{i1}+1] A^{i1}_{\a\b}\, v = - [p_{i1}-1] A^{1i}_{\a\b}\, v = 0
\quad \Rightarrow \quad
v^{(i)}_h = S^{i1}_{\a\b}\otimes{\bar S}^{\a\b}_{i1}\,v\ .
\lb{viS}
\ee
Now taking into account that $v_h^{(i)} = 0\,$ and
$S^{i1}_{\a\b}  = S^{1i}_{\a\b} \,,\ {\bar S}_{{i1}}^{\a\b} = {\bar S}_{{1i}}^{\a\b}\,$ we infer
\be
0 = S^{i1}_{\a\b}\otimes{\bar S}^{\a\b}_{i1}\,v =
S^{1i}_{\a\b}\otimes{\bar S}^{\a\b}_{1i}\,v \quad\Rightarrow\quad
w^{(i)}_h = A^{1i}_{\a\b}\otimes{\bar A}^{\a\b}_{1i}\,v\ .
\lb{wiA}
\ee
Using this property, we have been able to show "by brute force", in the case $n=3\,,$ that $w_h^{(2)} = 0\,$ for small values
of the level $k\,.$
Finding the appropriate combinatorial arguments in the general case (of arbitrary $n\,, i\,$ and $h$) remains a
challenge.

\begin{acknowledgement}

The authors thank Peter Dalakov, Joris Van der Jeugt, Ivan Penkov, Todor Popov, Ivan Todorov, Fernando Rodriguez Villegas and Paul Zinn-Justin for their interest and for sharing valuable information on the subject with them. LH would like to express his gratitude to Prof. V. Dobrev and the members of the local Organizing Committee of the 11th International Workshop "Lie Theory and Its Applications in Physics" (Varna, Bulgaria, 15-21 June 2015). The work of LH has been supported in part by Grant DFNI T02/6 of the Bulgarian National Science Foundation.

\end{acknowledgement}


\end{document}